\documentclass[12pt,preprint]{aastex}
\usepackage{emulateapj5}
\usepackage{xspace}

\newcommand{\chandra}{{\it Chandra}\xspace} 
\newcommand{\xmm}{{\it XMM-Newton}\xspace}

\newcommand{\rosat}{{\it ROSAT}\xspace}
\newcommand{\rxj}{RX\,J1856\xspace}
\newcommand\lax{\>\vcenter{\hbox{$<$\hskip-.75em\lower1.0ex\hbox{$\sim$}}}\>}
\newcommand\uax{\>\vcenter{\hbox{$>$\hskip-.75em\lower1.0ex\hbox{$\sim$}}}\>}

\begin{document}

\title{Isolated magnetar spin-down, soft X-ray emission and RX\,J1856.5-3754 } 
\author{Kaya Mori\altaffilmark{1} and Malvin A. Ruderman\altaffilmark{2}} 
\altaffiltext{1}{Columbia Astrophysics Laboratory, New York, NY 10027}
\altaffiltext{2}{Columbia University, Physics Department, New York, NY 10027}

\begin{abstract} 
When an isolated magnetar with magnetic dipole field $B\sim10^{15}$ G moves at high
velocity ($v > 10^7$ cm\,s$^{-1}$) through the ISM, its transition into
propeller 
effect driven spin-down may occur in less than $10^6$ years. We propose that
the nearby neutron star RX\,J1856.5-3754 is such a magnetar, and has spun down by the propeller
effect to a period greater than $10^4$ sec within $\sim5\times10^5$ years. This magnetar scenario is  consistent with
 observed thermal X-ray emission properties and the absence of detectable
spin-modulations of them.  Detection of other rapidly moving long period
($>100$ sec) magnetars with 
known ages would strongly
constrain the very great variety of predicted propeller-effect torque magnitudes.        

\end{abstract} 

\section{Introduction}

Isolated magnetars (spinning-down neutron stars with dipole fields
$10^{14}$--$10^{15}$ G) differ from canonical ($10^{11}$--$10^{13}$ G) pulsars
in several ways: 

(1) they spin-down very much more rapidly; 

(2) when relatively young (age $\lax 10^6$ years), their spin-down torque
    mechanism may change over from the relativistic particle plus magnetic
    field wind of 
    canonical pulsars to a ``propeller'' interaction \citep{davidson73,illarionov75} on a surrounding
     interstellar medium if they are rapidly moving through it; 

(3) their strongly magnetized surface can have very different emissivities for
    extraordinary and ordinary mode thermal X-rays. 

We consider below the extraordinarily varied spin-down descriptions for magnetars
when various proposed propeller interaction models are used
\citep{illarionov75, davies79}. Especially
significant may be a possible almost complete quenching of the spin of some magnetars within $10^6$ years,
not possible for isolated canonical pulsars or for magnetars embedded in
their own accretion disks. We  apply such considerations to understanding the observed X-ray emission
from the rapidly moving isolated $10^6$ year old neutron star RX\,J1856.5-3754
(hereafter \rxj). It has featureless nearly
blackbody thermal X-ray emission (with a paradoxically small blackbody radius)
which shows no indication of spin-period modulation ($< 1$ \%) in the range 10
ms $< P < 10^4$ sec \citep{ransom02, drake02, burwitz02}. 

Four possible explanations of the lack of detected spin-modulation of the
surface X-ray emission of an isolated $10^6$ year old NS are 

(1) Our line of sight to this pulsar is almost exactly along the pulsar
    spin-axis \citep{braje02}; 

(2) the magnetic field of this neutron star, unlike that of all known pulsars,
    is almost exactly axially symmetric with respect to the neutron stars
    spin-axis \citep{braje02}; 

(3) this particular neutron star was born less than $10^6$ years ago as a weak field    millisecond pulsar ($P<10$ ms) 

(4) this neutron star's spin has been almost quenched to $P>10^4$ sec. 
Of these (4) may be the most plausible. Whatever the details of the mechanism
for achieving it are, the needed effective spin-down torque at very long
periods seems to require both a transition out of the canonical pulsar
spin-down mechanism, where energy and angular momentum are lost from an almost
corotating magnetosphere at its very distant ($\lax 10^{14}$ cm) light
cylinder radius, and a huge surface dipole magnetic field. These two
requirements encourage exploring the possibility that \rxj is a magnetar,
rapidly moving through the ISM or a molecular cloud, is now spinning-down by propeller ejection of
that incident matter. 

The propeller effect may be an efficient spin-down mechanism when the
spin-period becomes larger than seconds, but proposed spin-down torques from
it are highly
model-dependent. We find that \rxj should spin-down to a period 
$>10^6$ sec for some  proposed propeller-effect models. The magnetar scenario is also consistent with 
observed spectral properties for both a single temperature and
two-component blackbody model. 

\section{Spin-down evolution}

When an isolated neutron star (NS) is spun-down by the torque of its relativistic wind
emission, 
\begin{equation} 
I\dot{\Omega} \sim -\frac{\mu^2 \Omega^3}{c^3} \equiv -K\Omega^n, 
\label{eq_canonical} 
\end{equation}  
where $I$ is the NS moment of inertia, $\mu$ is the NS magnetic dipole
 moment and $\Omega$ is the NS spin-velocity. The ``braking index'' n is 3 if
 $\mu$ and $\vec{\mu}\cdot\vec{\Omega}/\Omega$ are constant. When the NS is
 born with a spin-period much shorter than a presently observed one, the age of
 the star is the ``spin-down'' age  $\tau_{sd} \equiv \frac{P}{(n-1)\dot{P}}$,   where $P$ and $\dot{P}$ are the present spin period and its time derivative. 
The canonical spin-down age, $\tau_{csd} \equiv P/2\dot{P}$, differs from the
true age for $n\ne3$. If $n < 1$ $\tau_{sd}$ clearly must have a very different
meaning.  
When $n > 1$, the canonical spin-down age indicates {\it the time since the NS
entered the spin-down phase described by equation (\ref{eq_canonical})}: 
\begin{equation}
\tau_{sd} = \frac{2}{n-1} \tau_{csd},  
\label{eq_csd1}
\end{equation}
as long as the original spin-period $\ll P$ but when $n < 1$, the canonical spin-down age indicates {\it the approximate time
remaining before the 
NS spin will be completely quenched ($\tau_f$):} 
\begin{equation}
\tau_{f} = \frac{2}{1-n} \tau_{csd}.   
\end{equation} 
To quench a NS's spin, that spin-down must then be in a phase described by
$n<1$ in equation (\ref{eq_canonical}). 

When the NS light cylinder radius ($c/\Omega$) becomes larger than the radius
of its corotating magnetosphere ($R_m$), the spin-down mechanism is changed to
one in which the spin-down torque is from direct spin-up of the surrounding
medium (``propeller effect''). In this case, the braking index is no longer
$\sim3$. In proposed models, it is generally in the
range $-1\le n \le 2$. 

Propeller effect spin-down torque depends upon three parameters, the NS
magnetic dipole moment ($\mu$), the density of incoming gas at the
magnetosphere boundary with the surrounding medium ($\rho_m$)  and velocity of the incoming gas at the
magnetosphere boundary in the NS rest frame ($v_m$).  Assuming the NS
spin-down torque has fixed power-law dependences on these parameters,
dimensional analysis gives    
\begin{equation} 
I\dot{\Omega} = - \kappa\, \mu^{(3+n)/3} \rho_m^{(3-n)/6} v_m^{(3-4n)/3}
\Omega^n       
\end{equation}  
with $\kappa$ and $n$ dimensionless constants. 
In the following  discussion, we set NS mass and radius to  $1.4M_{\odot}$ and
10 km respectively.

\subsection{Models for propeller caused spin-down} 

Two parameters can classify the many proposed propeller spin-down
models. In these models, spin-down torque is    
\begin{equation} 
I\dot{\Omega}=-\dot{M} R_m v_m \mathcal{M}^{\gamma},   
\label{eq_sd}
\end{equation} 
where $\dot{M} = \rho_m (\pi R_m^2) v_m$ is the incoming (and largely ejected)
mass rate. $\mathcal{M}$ is the Mach number defined as the ratio of incoming
medium velocity to NS spin-velocity at the magnetosphere boundary: $\mathcal{M}
\equiv R_m\Omega/v_m$. Proposed propeller models have $\gamma$ between $-1$ and
$2$. $\mathcal{M}>1$ and $\mathcal{M}<1$ are the supersonic and subsonic
regimes, which have been treated differently in proposed models. 

The ram pressure balance at the (non-spherical) magnetosphere boundary from
either $v_m$ or $\Omega R_m$ is assumed to be   
\begin{equation} 
\frac{B_m^2}{8\pi} = \rho_m v_m^2 \mathcal{M}^{\delta}, 
\label{eq_ram} 
\end{equation} 
where $B_m$ is local magnetic field strength at the magnetosphere boundary. Since $B_m =
 \mu/R_m^3$ for a dipole magnetic field with dipole moment $\mu$, the braking index $n$ is then 
\begin{equation} 
n=\frac{3(2\gamma-\delta)}{\delta+6}. 
\end{equation}  
The spin-down torque is 
\begin{equation}
I\dot{\Omega}=-\frac{\mu^2}{R_m^3}\mathcal{M}^{\gamma-\delta}.  
\label{eq_torque}
\end{equation}

A remarkably large number of different propeller spin-down models have been
proposed and applied. The braking indices suggested by these propeller models are in the range of $-1 \le n \le 2$. Some of the models were intended only
for NSs with spin-down from surrounding accretion disks. 
Table \ref{tab_models} categorizes most of the proposed propeller effect
models by their $\gamma$ and $\delta$.   

We note that at the transition into propeller spin-down of an isolated pulsar,
$\Omega R_m \sim c$ so that $\mathcal{M} \sim 10^3$ in equations (\ref{eq_sd}),
(\ref{eq_ram}) and (\ref{eq_torque}). In the models of Table
\ref{tab_models} (by various authors, and,  sometimes, the same authors at
different time) the $\gamma$--$\delta$ exponent in equation (\ref{eq_torque})
varies from $-1$ to $2$ corresponding to differences of $10^9$ in the magnitude of
the propeller driven spin-down torques which take over from that of
equation (\ref{eq_canonical}).    

In most models $\gamma\ne\delta$ so that  in the transition from pulsar to
propeller phase there is a huge discontinuity in spin-down torque. We favor a
plausible propeller spin-down model with $\gamma=\delta$ which gives no such
discontinuity.  

A particularly simple model completely ignores all magnetic
fields within both the incoming and ejected ionized beams and also possible
collisionless collection stream-stream interactions. Because both
flows are so very dilute ($n_m \lax 10$ cm$^{-3}$) single particle interaction mean free
paths $\gg R_m (\sim 10^{12}$ cm). Incidence and reflection of the ISM on the
stellar magnetosphere could then be described as a sum of single particle
trajectories. A "rough" interface
model (RIM) is one in which the interface between the corotating
magnetosphere
of the NS and the ISM flow onto it is
approximated as a sawtooth-like  structure in the corotating frame, with
angles between the sawtooth surfaces and the surface of constant
radius from the star  assumed to be of order a radian. In  "smooth"
interface models that interface is assumed to be much more spherical
(typically with angular deviations of order the inverse Mach number of
the very supersonic propeller motion).  In a "rough" interface model the 
deflected incident flow pushes strongly with a comparable force in
the tangential and radial directions. This gives $\gamma=\delta=1$ and
$n=3/7$. (In a smooth interface limit, $\gamma=-1, \delta=0$ gives the $n=-1$
of \citet{illarionov75}). In our discussions below we give
particular emphasis to it.  

\subsection{Transitions between different spin-down stages of magnetars} 

We discuss three transitions in the spin evolution of magnetars as $\Omega$
decreases. When $\Omega =
c/R_m$ there is a transition from canonical pulsar spin-down 
to propeller spin-down. When $\Omega = v_m/R_m$, there is the transition from supersonic
($\mathcal{M}>1$) to subsonic propeller spin-down ($\mathcal{M}<1$). When
$\Omega = \Omega_K(R_m)$ (the Keplerian angular velocity at the
magnetosphere boundary), there is a transition from propeller spin-down
to strong accretion. Table \ref{tab_trans} summarizes spin-periods and
transition times for different propeller spin-down models. 

\section{The exceptional isolated neutron star \rxj \label{sec_1856}}

The isolated neutron star \rxj was serendipitously discovered by \rosat
as an X-ray source with thermal emission \citep{walter96}. Detection of  
proper motion provided a distance $\sim$ 120 pc and velocity
$\sim$ 200 km\,s$^{-1}$ and supports an association of \rxj with the Upper
Scorpius birthplace indicating  a NS age $\sim5\times10^5$ yrs \citep{walter02, kaplan02}. Deep \chandra and \xmm observations with total exposure time $\sim$ 500 ks were
performed to search for spectral feature and pulsation. However, neither was
found in the X-ray data which had high resolution and sensitivity
\citep{ransom02, drake02}.    
   
This lack of observed pulsation is perplexing since 
modulation of an anisotropic temperature distribution suggested by the
two-component blackbody model fit discussed below seems expected, and pulsation
modulation of X-ray emission has been detected in 
other isolated NSs with thermal emission \citep{haberl97, zavlin00}.
We propose that \rxj has been spun-down to a spin period longer
than $10^4$ sec by the propeller effect. The fraction of magnetars among
isolated NSs is $\sim 10^{-1}$ \citep{kouveliotou98}, and is much larger among NSs from which thermal
emission has been detected. We find that the only plausible way to achieve the
needed condition for early transition
into the propeller phase and rapid enough spin-down thereafter is for this star to be
a magnetar. 

We first assume $B\ge 10^{14}$ G, sufficiently large that transition into the
propeller phase occurs in much less than $5\times10^5$ years. Then, we
searched parameter space in the $\gamma$--$\delta$ plane for the region in
which \rxj could, thereafter, have been spun-down to a period longer than $10^4$ sec. Input
parameters are magnetic dipole moment $\mu$, incoming gas density $n_m$, and
incoming gas velocity $v_m$. We fixed $v_m$ from the observed stellar proper
motion at 200 km\,$^{-1}$. 

ISM particles (mainly hydrogen) at the magnetosphere radius will be charged
and be reflected by the stellar magnetosphere. [Thermal photons from the NS
surface ionize all hydrogen in the vicinity of the star. The
ionization time of hydrogen at $R_m\sim10^{12}$ cm ($\sim10^2$ s) is much shorter than $R_m/v_m$ ($\sim 10^5$ s). Therefore, hydrogen is fully-ionized at the
magnetosphere boundary before the star reaches that region].    
 
\begin{figure}[h]
\epsscale{1.0}
\plottwo{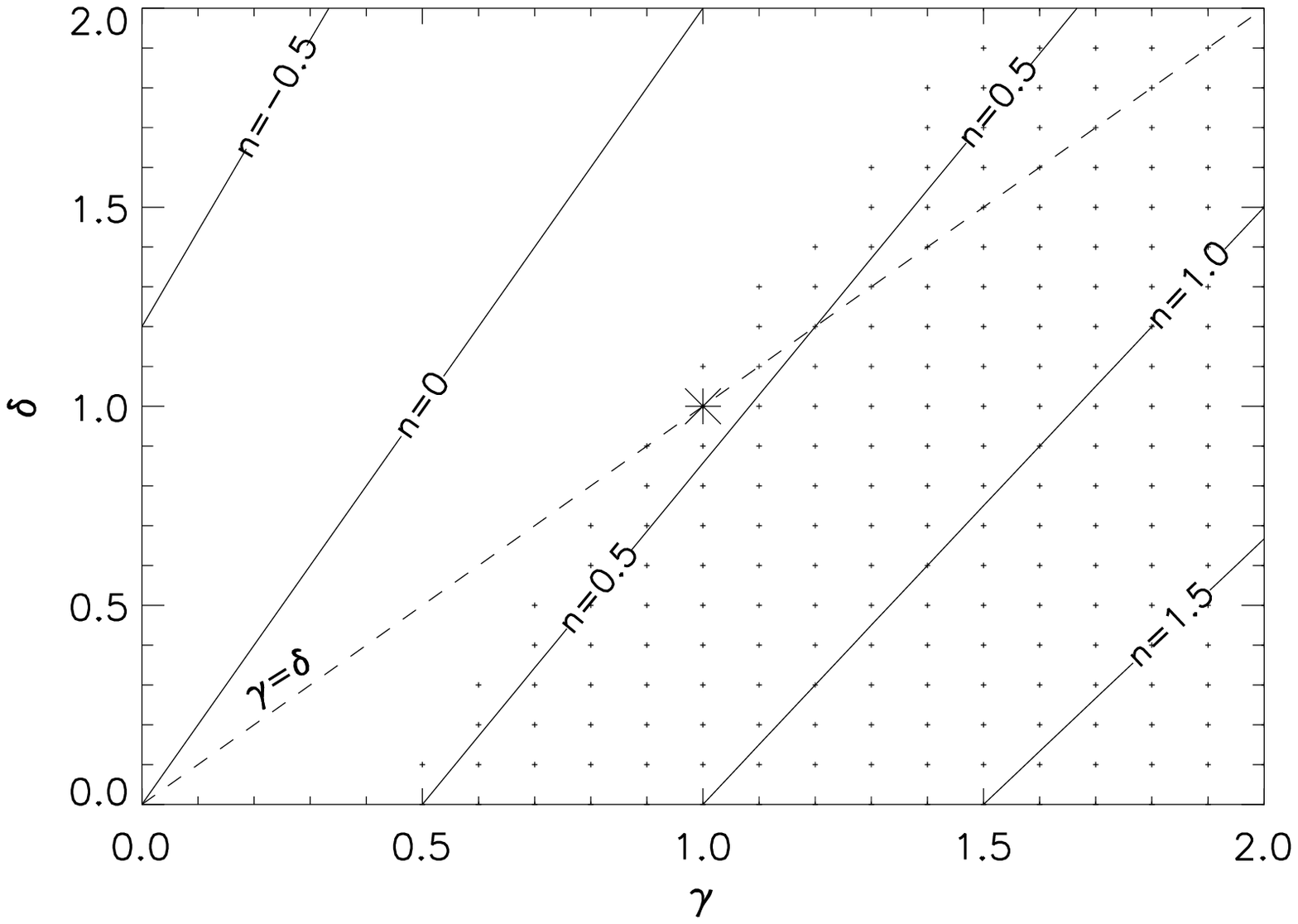}{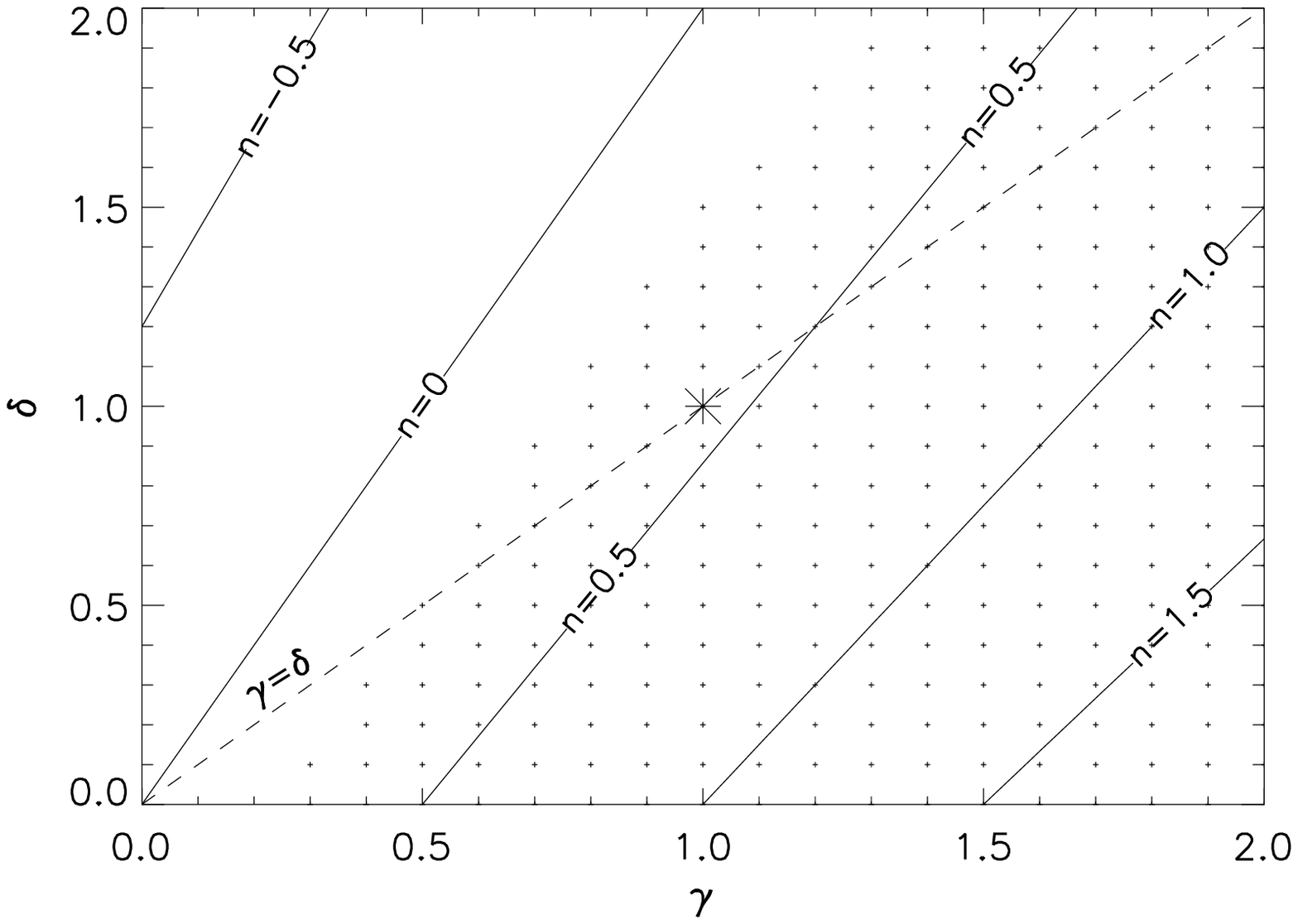}
\caption{$\gamma$--$\delta$ diagram for $\mu_{33}=5$ and $n_m=1$ cm$^{-3}$ (left) and
$\mu_{33}=0.5$ and $n_m=10^5$ cm$^{-3}$ (right). The dotted region shows he
$\gamma$--$\delta$ parameter space for which \rxj can spin-dwon to $P>10^4$
sec within $5\times10^5$ years. The asterisk refers to our
favorite model ($\gamma=\delta=1$). The ISM density in the vicinity
of \rxj has been  
estimated to be $\sim 1$ cm$^{-3}$
from studies of the nearby H$\alpha$ nebula \citep{vankerkwijk01_2}. \rxj may
have passed through the nearby molecular cloud R CrA with 
density $>10^5$ cm$^{-3}$ \citep{giannini98}. \label{fig_diagram}} 
\end{figure}

Figure \ref{fig_diagram} shows the $\gamma$--$\delta$ parameter space for which
\rxj can spin-down to $P > 10^4$ sec within $5\times10^5$ years. Our rough
interface model ($\gamma=\delta=1, n=3/7$) can achieve the required spin-down
for $B\sim5\times10^{15}$ G ($\mu_{33}=5$) and $n_m=1$ cm$^{-3}$ and 
$B\sim5\times10^{14}$ G ($\mu_{33}=0.5$) and $n_m=10^5$ cm$^{-3}$
(estimated density of the nearby molecular cloud R CrA through which \rxj may 
have passed \citep{giannini98}). 

\subsection{Thermal emission from \rxj} 

The  surface temperature of \rxj is roughly consistent with predictions of standard cooling curves
at its age $\sim5\times10^5$ years \citep{tsuruta02}. \rxj may radiate because
of accretion at an incoming mass rate $\dot{M}=\pi
R_m^2 v_0 \rho_m = 5\times10^7R_{m,12}^2 v_{m,7} n_m$ g\,s$^{-1}$. However, the
X-ray luminosity from such an accretion rate is below the 
detection limit of \chandra and \xmm observations \citep{rutledge01_2}. A
third heat source may be the continuous dissipation of the large magnetic
field energy in the stellar crust \citep{heyl98}. 

Given featureless X-ray spectra, several approaches using spectral
energy  distribution (SED) in the optical and X-ray band have been undertaken to reveal the surface composition \citep{pons02}. Light element atmospheres (H
 and  He) were ruled out since they predict about two orders of magnitudes larger optical flux compared to the  
observed values \citep{pons02}. On the other hand, a blackbody model
 underpredicts optical flux by a factor of $\sim 7$  \citep{walter02}. Non-magnetized heavy element
 atmosphere models (e.g. Si-ash or Iron) predict correct SED over the optical
 and X-ray band  with a single temperature ($kT^{\infty} = 40$ eV) \citep{walter02}. However, they have absorption features which deviate 
 from the featureless X-ray data \citep{burwitz02}. A similar situation exists for
 magnetized heavy element atmospheres \citep{rajagopal97}.  

A two-component blackbody model was proposed to account for the
multi-wavelength SED and the featureless X-ray spectra \citep{pons02}. Hot and
cold blackbody components for X-ray and optical spectra respectively are
consistent with both SED and featureless X-ray data \citep{pons02, braje02}. However, it is hard to
obtain blackbody-like spectra when significant atmosphere is present on the
surface. 

At sufficiently high magnetic field strength and low temperature, a NS
surface becomes very dense liquid or solid with almost no atmosphere above it
\citep{ruderman71, lai96}. 

That \rxj is a magnetar is consistent with the two temperature
model (anisotropy of surface temperature distribution caused by strong
magnetic field) and the model for a condensed iron surface because they both
require strong magnetic field strength on the surface. The emissivity of 
such a condensed matter surface in $10^{15}$ G is about 1/2 that of a
blackbody because the strongly magnetized very dense ($\rho\sim
560Z^{-3/5}B_{12}^{6/5}$ g\,cm$^{-3}$, with $B=10^{12}B_{12}$ G) has essentially no emissivity for O-mode X-rays
but is near blackbody for E-mode ones. Then the average emissivity for the
radiating surface is $\sim$ 30--50\% depending on photon energy and magnetic field
geometry \citep{zane03}. The actual NS area should then be more than twice that of
the apparent blackbody area so that the inferred NS radius becomes 
$\uax \sqrt{2} R_{BB}$ \citep{zane03}. 

\section{Discussion}

Will we find other isolated NSs in the propeller phase? 
There are several NSs with discrepant supernova remnant and ``canonical''
spin-down ages ($\tau_{csd}$). In some cases, the 
discrepancy may be due to the fact that NSs are in a propeller
phase. (Alternatively, they could have been born with a long spin period close
to present value.) However, some of them have spin periods shorter than $< 1$
sec, so it is difficult for them to enter a propeller phase unless the ambient
gas density is very large ($n_m \gg 1$ cm$^{-3}$). 

Detection of long pulsation periods ($> 10$ sec) from isolated NSs may be
another indication of NSs in a propeller phase. Potential candidates for such isolated
NSs would have ages between $10^3$ years (enough time for spinning down to
enter the propeller phase) and $10^6$ years (still detectable thermal
emission). In a magnetar, magnetic field decay processes can keep the magnetar
more X-ray luminous than would be the case for canonical pulsars \citep{heyl98}. Then older magnetars might be still observable by
their X-ray emission after canonical NSs have faded. 

Detection of spin-period variation in X-ray flux over several X-ray
observations would explore periods $>10^4$ sec for the case of \rxj. In most cases
in which \rxj has spun-down to $P>10^4$ sec (\S \ref{sec_1856}), it should at present have a period of at least $10^6$ sec. The chance of 
seeing a different surface aspect of the star may be small and gravitational 
light-bending effects will smear out some spectral modulation. 

If an isolated NS in a propeller phase is confirmed it will select among
propeller effect models. If a NS is in the propeller phase, one should be cautious in
deriving NS ages and magnetic field strengths from $P$ and $\dot{P}$
measurements since the conventional dipole radiation wind formulae of equation
(\ref{eq_canonical}) and (\ref{eq_csd1}) are no longer
valid. In such cases, independent determination of ages and
magnetic field  strengths (e.g. supernova remnant ages and spectroscopic
measurements of B-field) are needed. 

\acknowledgments{We thank J.J. Drake, R.V.E. Lovelace, J. Pringle,
J.E. Tr{\"u}mper, M.H. van Kerkwijk and S. Zane for useful discussions. }

\begin{deluxetable}{cccccccc}
\tablewidth{0pt}
\tablecaption{Propeller effect model parameters. \label{tab_models}}
\tablehead{\colhead{$\gamma$} & \colhead{$\delta$} &
\colhead{$n$} &         & \colhead{$n_\mu$}& \colhead{$n_\rho$} & \colhead{$n_v$}       & \colhead{Models}} 
\startdata
$-1$       & 0  & $-1$  && 2/3  & 2/3   & 7/3   & \citet{illarionov75}\\
0       & 0     & 0     && 1    & 1/2   & 1     &\citet{davidson73}\\
1       & 0     & 1     && 4/3  & 1/3   & -1/3  &\citet{menou99}    \\
2       & 0     & 2     && 5/3  & 1/6   & -5/3  & \citet{davies79}\\
1       & 1     & 3/7   && 8/7  & 3/7   & 3/7   & A rough interface model (RIM)\\
2      & 2     & 3/4   && 5/4  & 3/8   & 0     &  \citet{romanova02}\tablenotemark{a}   \\

\enddata
\tablenotetext{a}{Their analytic model gives $n=3/5$ which differs from the $n=3/4$ because they used the
outflow density which depends on $\Omega R_m$ (see more details in section 2
of \citet{romanova02}). In their numerical model, $n=1.3$ and $n_{\mu}=0.8$. }
\end{deluxetable}

\begin{deluxetable}{lcc}
\tablewidth{0pt}
\tablecaption{ NS periods and transition ages for different pulsar spin-down
mechanisms. \label{tab_trans}}
\tablehead{\colhead{Transition} & \colhead{Period [sec]} &
\colhead{Transition time [yrs]\tablenotemark{a}}} 
\startdata
Pulsar wind $\rightarrow$ Propeller & $37 \mu_{33}^{1/3} n_m^{-1/6} \beta_m^{(\delta-2)/6}$	&$2.2\times10^4 \mu_{33}^{-4/3} n_m^{-1/3} \beta_m^{(\delta-2)/3}$\\
Supersonic $\rightarrow$ Subsonic & $1.6\times10^6 \mu_{33}^{1/3} n_m^{-1/6}
v_{m,7}^{-4/3}$ & [$10^5$--$10^7$]\tablenotemark{b} \\
Propeller $\rightarrow$ Accretion & $\sim [10^6]$\tablenotemark{b} & [$10^5$--$10^7$]\tablenotemark{b} 
\enddata
\tablecomments{$\beta_m \equiv v_m/c$ and $\mu=10^{33}\mu_{33}$
G\,cm$^{3}$. $\mu_{33}=1$ corresponds to $B=10^{15}$ G and $R=10$ km.}
\tablenotetext{a}{Age since birth.} 
\tablenotetext{b}{These do not usually have simple power-law forms and we show
approximate ranges. We assumed $\gamma=\delta=1$. The parameter ranges we considered here are:
$B\sim10^{14}$--$10^{15}$ G, $n_m\sim 1$--10 cm$^{-3}$ and $v_m\sim 10$--$100$
km\,s$^{-1}$.} 
\end{deluxetable}

\end{document}